\documentclass[a4paper,fleqn,usenatbib]{mnras}

\usepackage[T1]{fontenc}
\usepackage{ae,aecompl}

\usepackage{graphicx}	
\usepackage{amsmath}	
\usepackage{amssymb}	

\title[Cosmic ray spectral hardening]{Non-linear acceleration at supernova remnant shocks and the hardening in the cosmic ray spectrum.}

\author[S. Recchia, S. Gabici]{
	S. Recchia,$^1$\thanks{E-mail: recchia@apc.in2p3.fr}		
	S. Gabici,$^{1}$\thanks{E-mail: gabici@apc.in2p3.fr}\\
	{$^1$ \it APC, Universit\'e Paris Diderot, CNRS/IN2P3, CEA/Irfu, Observatoire de Paris, Sorbonne Paris Cit\'e, France.}
}

\date{Accepted XXX. Received 2016 April 26; in original form 2016 April 26}

\pubyear{2017}

\begin{document}
\label{firstpage}
\pagerange{\pageref{firstpage}--\pageref{lastpage}}
\maketitle

\begin{abstract} 
In the last few years several experiments  have shown that the cosmic ray spectrum below the knee is not a perfect power-law. In particular, the proton and helium spectra show a spectral hardening by $\sim 0.1-0.2$ in spectral index at particle energies of $\sim \rm 200-300\, GeV$/nucleon.  Moreover, the helium spectrum is found to be harder than that of protons by $\sim 0.1$ and some evidence for a similar hardening was also found in the spectra of heavier elements.
Here we consider the possibility that the hardening may be the result of a dispersion in the slope of the spectrum of cosmic rays accelerated at supernova remnant shocks. 
Such a dispersion is indeed expected within the framework of non-linear theories of diffusive shock acceleration, which predict steeper (harder) particle spectra for larger (smaller) cosmic ray acceleration efficiencies.

\end{abstract}



\section{Introduction}
\label{sec:intro}
In the standard picture of the origin of cosmic rays (CR) the observed flux, at least below the energy of the "knee" ($\rm E_{knee} \approx 3\, PeV$; see e.g \citealt{Hoorandel-2006}, is thought to be produced in the Galactic disc at supernova remnant (SNR) shocks through diffusive shock acceleration (DSA) (\cite{Drury-1983RPPh...46..973D}). After leaving their sources, CRs are believed to propagate diffusively through the interstellar medium (ISM), and eventually escape from the Galaxy (see e.g \citealt{Berezinskii90}). The observed CR spectrum below the knee resembles  a power law in energy $\approx \rm E^{-2.7}$ and can be roughly accounted for if one assumes that both the slope of the injection spectrum and the CR diffusion coefficient are power laws in energy, with slopes $-\gamma$ and $\delta$ respectively. Various observational constraints provide $\gamma + \delta \approx 2.7$ and $\delta \approx 0.3-0.6$ (see e.g \citealt{Strong-1998ApJ...509..212S}; \citealt{Evoli-2008JCAP...10..018E}; \citealt{Blasi-Amato-2012JCAP...01..010B}).

However, in the last few years numerous evidences have been collected which point to a more complex scenario: most notably several experiments such as ATIC-2 (\citealt{Atic2-2009}), PAMELA (\citealt{Adriani2011}), CREAM (\citealt{Cream-2011}) and AMS-02 (\citealt{Aguilar-2015}) found a spectral hardening in the proton and helium spectra at particle energies $\rm 200-300\, GeV$/nucleon. Moreover, the helium spectrum is found to be harder than that of protons by $\sim 0.1$ in spectral index. Some evidence for a similar hardening was also found in the spectra of heavier elements (\citealt{Maestro-2010}).
The PAMELA data (\citealt{Adriani2011}) suggest that the slope of protons changes from $\gamma_1 \sim 2.85$ (below $\sim 230 \rm \, GeV $) to $\gamma_2 \sim  2.67 $ (above $\sim 230 \rm \, GeV$), with a slope change of $\sim 0.18$. Instead, the AMS-02 data found the break at $\sim 335 \rm \, GeV$ and a slope change of $\sim 0.13$.

This spectral feature is still not understood and several explanations have been put forward in which the spectral hardening is interpreted as: the result of a break in the CR diffusion coefficient  (\citealt{Tomassetti}; \citealt{Genolini-2017}); a consequence of  the transition from the scattering of CRs  on  self-generated waves to scattering on pre-existing waves (\citealt{Blasi-2013JCAP}); the effect of a nearby source (\citealt{Thoudam-2012-MNRAS}); the consequence of a  dispersion in spectral index at the sources (\citealt{Yuan-2011}); the result of non linear effects in DSA at SNRs (\citealt{Ptuskin-2013-pamela}); the possible presence of distinct populations of CR sources (\citealt{Zatsepin-2006});  the consequence of a break in the energy loss rate (\citealt{Krakau-2015}).\\

In the present paper we suggest that the spectral hardening may be a natural prediction of the non-linear theory of diffusive shock acceleration.
It would result from the interplay of the efficient magnetic field amplification at SNR shocks and of the  CR Alfv\'{e}nic drift  in the upstream region.
Following \cite{Caprioli-2012JCAP...07..038C}, we will show that these two effects may result in a dispersion in the CR spectral slope at the sources which may lead (see also \citealt{Yuan-2011}) to the observed hardening. 
The reasons for considering such scenario are manifold: first of all, the presence of efficient magnetic field amplification has been detected in several SNRs (see e.g \citealt{Ballet-2006}; \citealt{Vink-2012}) and is widely considered  as a crucial ingredient for the acceleration  of CRs to the energy of the knee (see e.g \cite{Bell-2004MNRAS.353..550B}).  Second, recent observations of $\gamma$-rays in  SNRs both in the GeV (see e.g \citealt{Fermi-2011ApJ...734...28A}) and in the TeV band (see e.g \citealt{Acciari-2011}) show that there may be a quite large dispersion in the slope of the CR spectrum at SNRs. In fact, in the cases in which the observed $\gamma$-ray emission is likely of hadronic origin (see e.g \citealt{Morlino-2012A&A...538A..81M}), the inferred CR spectrum shows a quite large dispersion in the  spectral index, $\rm \propto E^{-2.1}- E^{-2.5} $  (see e.g \citealt{Caprioli-2011JCAP...05..026C}; \citealt{Yuan-2011} and references therein).\\

It is interesting to note that such  spectra are significantly steeper than the universal spectrum $\rm \propto E^{-2.0}$ predicted by the {\it linear} (test-particle) theory of DSA at SNR shocks. 
At first sight, the disagreement seems to be even larger if one considers {\it non-linear} theories of DSA (NLDSA), which account for the reaction of CRs on the shock dynamics (see e.g \citealt{Drury-1983RPPh...46..973D}; \citealt{Blandford-1987PhR...154....1B}; \citealt{Jones-1991SSRv...58..259J}; \citealt{Blasi-2002}).
In this context, the pressure exerted by CRs onto the upstream fluid induces the formation of a shock precursor, which in turn makes the CR spectrum at the shock concave, namely, steeper than  $\rm \propto E^{-2.0}$ at low energies and harder at high energies. Moreover, the more efficient the CR acceleration the more evident becomes the concavity and for large efficiencies  the  spectrum above few GeV becomes as flat as $\rm \propto E^{-1.5} $, clearly at odd with the gamma-ray observations of SNRs reported above.

For this reason, in the last few years a number of works have been devoted to the study of possible ways to reconcile the predictions of NLDSA theories with observations. In particular it has been proposed that taking into account the velocity of the CR scattering centers (see e.g \citealt{Ptuskin-2008AIPC.1085..336Z}; \citealt{Caprioli-2012JCAP...07..038C}) and the effect of the amplified magnetic field at the shock (see e.g \citealt{Vainio-1999A&A...343..303V}; \citealt{Caprioli-2012JCAP...07..038C}) the CR source spectrum may become significantly steeper than $\rm \propto E^{-2.0}$. In this paper we focus on the work by
\citealt{Caprioli-2012JCAP...07..038C}, where a theory of NLDSA is developed, which includes both the magnetic field amplification, due to CR streaming instability, and the resulting enhanced velocity of the CR scattering centers (Alfv\'en waves propagate faster for larger values of the ambient magnetic field (see e.g. \citealt{Ptuskin-2008AIPC.1085..336Z}, \citealt{Caprioli-2012JCAP...07..038C}).

The main findings of \cite{Caprioli-2012JCAP...07..038C} can be summarised as follows: 
\begin{enumerate}
\item{the magnetic field amplification acts as a self-regulating mechanism of the acceleration process. The pressure of the amplified magnetic field makes the shock compression factor  smaller than 4 (which is the strong shocks limit of the test particle regime of DSA, where the field is not amplified). As a consequence of that, the maximum acceleration efficiency for strong shocks turns out to be $\sim 30\%$ and the shock modification induced by the CR pressure is quite modest. This is quite at odd with earlier formulations of NLDSA, in which the CR acceleration efficiency can reach values well above $\sim 30\%$ and the compression factor can be well above 4 (see e.g. \citealt{Blasi-2005MNRAS.361..907B})}
\item{the spectrum at energies above few GeV resembles quite well a power law whose  spectral slope remains virtually constant, together with the acceleration efficiency, for a large part of the SNR lifetime (up to $\sim$ 20000-30000 yr);}
\item{the combined effect of the magnetic field amplification and of the  Alfv\'{e}nic drift in the upstream region makes the  spectrum  steeper than $E^{-2.0}$, with slopes in the range  $2.1-2.6$ in agreement with the $\gamma$-ray observations of SNRs. In addition to that, the more efficient is the CR acceleration the steeper is the spectrum, contrary to the standard predictions of NLDSA, in which an efficient acceleration leads to harder spectra.}
\end{enumerate}
For a more extended discussion on this approach and its limits of validity, the reader is referred to \citet{Caprioli-2012JCAP...07..038C}.

Based on the results summarized above, in the following we treat the acceleration efficiency at SNR shocks as a free parameter in the range $\xi_{CR} \approx 0.03-0.3$, and we assume that the CR spectrum at SNR shock is a power law of slope $\gamma_{CR}$. 
We then compute the spectral slope as a function of the acceleration efficiency $\gamma_{CR}(\xi_{CR})$ by taking into account the magnetic field amplification by CR streaming instability and the effect of the velocity of the self-generated Alfv\'{e}n waves, which act as scattering centers.
In agreement with  \cite{Caprioli-2012JCAP...07..038C}, we find that  the dispersion in the acceleration efficiency induces a dispersion  in the spectral slope, with steeper (softer) spectra corresponding to larger (lower) acceleration efficiencies. 
Finally, taking into account  the dispersion in the spectral slope and the relation between acceleration efficiency and slope, the observed proton and helium spectral hardening  at $200-300$ GeV can be accounted for in a quite natural way.


The paper is organized as follows: in Section~\ref{sec:method} we illustrate the calculation of the compression factor felt by CRs as a function of the CR acceleration efficiency and we show the resulting dispersion in slope. In Section~\ref{sec:results} we use  these results to estimate the observed proton and helium spectrum with the additional assumption that CRs propagate diffusively in the Galaxy with a  power law diffusion coefficient (whose slope and normalization is chosen in order to  fit the data) and we compare the obtained spectrum with the data. Finally, we conclude in Section~\ref{sec:conc}. 

\section{Magnetic field amplification,  Alfv\'en speed and the proton spectral slope}
\label{sec:method}
Keeping in mind the results by \cite{Caprioli-2012JCAP...07..038C} on the  spectrum of accelerated particles at SNRs in the presence of magnetic field amplification by CR streaming instability and of Alfv\'{e}nic drift, in this section we illustrate a simple  calculation which allows to quickly estimate the  slope of the CR spectral slope under the following assumptions: 
\begin{enumerate}
\item{the CR acceleration efficiency $\xi_{CR}$ is an input parameter of the problem, and is  in the range $\xi_{CR} \sim 0.03-0.3$. Thus the shock modification is modest and the CR  spectrum is nearly a perfect power law, as found by \cite{Caprioli-2012JCAP...07..038C};} 
\item{the magnetic field is amplified by the CR streaming instability and is assumed to be the same in the whole upstream region;}
\item{the Alfv\'{e}n waves excited in the upstream propagate against the fluid at velocity $v_{A1}$, which is computed in the amplified magnetic filed, while they are assumed to be isotropized in the downstream  region, giving $v_{A2}\sim 0$ (the subscripts 1 and 2 refer to quantities calculated in the upstream and downstream region, respectively).}
\end{enumerate}

The slope of the CR spectrum depends on the effective compression factor felt by CRs, which in turn depends on the CR acceleration efficiency $\xi_{CR}\equiv P_{CR}/\rho_1u_1^2$, on  the fluid  and Alfv\'{e}nic Mach numbers upstream ($M_1^2 \equiv \rho_1u_1^2 / \gamma P_g$ and $M_A^2 \equiv \rho_1u_1^2 / 2P_w$ respectively) and on the jump conditions at the shock, where $\rho_1$, $u_1$, and $P_g$ are the upstream gas density, velocity and pressure, $P_{CR}$ is the CR pressure at the shock and $P_w$ is the pressure of the amplified (upstream) magnetic filed. 
Here, we neglect the modest shock modification induced by the CR pressure in the upstream fluid, which implies that all the relevant physical quantities characterising the fluid do not depend on the location upstream. Finally, we assume that the gas can be described by an adiabatic equation of state.\\

\begin{figure}
	\centering
	\includegraphics[width=\linewidth]{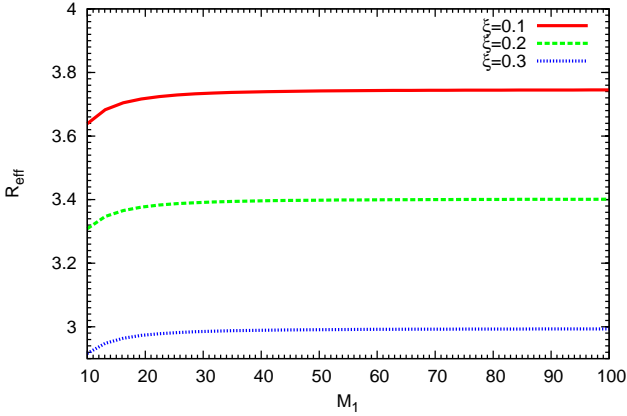}
	\caption{Effective compression factor felt by CRs at the shock as a function of the upstream fluid Mach number $M_1$ and for three different values of the CR acceleration efficiency: $\xi_{CR}=$ 0.1, 0.2 and 0.3.}\label{fig:Reff-M0}
\end{figure}

Following the calculations presented in \citet{Caprioli-2012JCAP...07..038C}), one can evaluate the effect of the magnetic field amplification due to CR streaming instability, and estimate the shock Alfvenic Mach number as a function of the CR acceleration efficiency (Equation 2.22 in \citealt{Caprioli-2012JCAP...07..038C}):
\begin{equation}\label{eq:MA}
M_A^2 = \frac{4}{25}\frac{\left[1-(1-\xi_{CR})^\frac{5}{4}\right]^2}{(1-\xi_{CR})^\frac{3}{4}}.
\end{equation} 
Another crucial parameter is the fluid compression factor $R = u_1/u_2$, which can be computed after taking into account the pressure of the amplified magnetic field. Following a procedure similar to that presented in \citet{Vainio-1999A&A...343..303V} and \citealt{Caprioli-2012JCAP...07..038C} we get:
\begin{align}
&\frac{M_1^2}{2}\frac{\frac{\gamma +1}{R} -(\gamma-1)}{1+\Lambda_B} \approx 1\label{eq:R}, \qquad \rm where \\
&\Lambda_B = W\left[1+R\left(\frac{2}{\gamma} -1\right)\right] \rm \qquad and \nonumber \\
& W=\frac{\gamma}{2}\frac{M_1^2}{M_A^2}. \nonumber
\end{align}

The effective compression felt by CRs differs from $R$, because in the upstream region the Alfvén waves generated by the CR streaming instability propagate in the direction opposite to the fluid. Since CRs are coupled to waves through scattering, the effective advection velocity they experience in the upstream region is not $u_1$, but rather $u_1-v_{A1}$.
Thus, the effective compression factor felt by CRs is
\begin{equation}\label{eq:Reff}
R_{eff}= \frac{u_1 -v_{A1}}{u_2} = R\left(1-\frac{1}{M_A}\right).
\end{equation} 
The CR spectral slope can then be estimated as (see e.g \citealt{Blandford-1987PhR...154....1B}; \citealt{Berezinskii90})
\begin{equation}\label{eq:gamma-CR}
\gamma_{CR}\sim \frac{3\,R_{eff}}{R_{eff}-1}.
\end{equation}
It is important to stress that only two physical parameters regulate the system: the fluid upstream Mach number $M_1$ and the CR acceleration efficiency $\xi_{CR}$.

In  Figure~{\ref{fig:Reff-M0}} we show the dependence of the effective compression ratio $R_{eff}$ on the Mach number $M_1$ for three different values of $\xi_{CR}$. For $M_1 \gtrsim 10$ the effective compression ratio (the same result also holds for the fluid compression ratio and for the spectral slope) is virtually independent on $M_1$. This implies that the slope of the spectrum of accelerated particles does not change for most of the SNR lifetime.
This is in agreement with the findings of \citet{Caprioli-2012JCAP...07..038C}. 

\begin{figure}
	\centering
	\includegraphics[width=\linewidth]{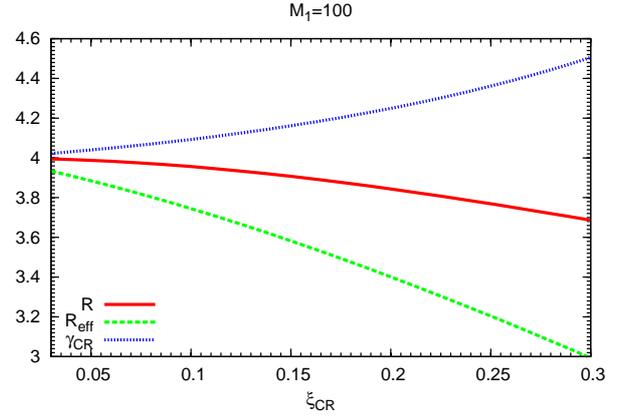}
	\caption{Fluid compression factor $R$, effective compression factor $R_{eff}$ and CR spectral slope $\gamma_{CR}$ as a function of the CR acceleration efficiency $\xi_{CR}$. The value of the fluid Mach number is $M_1 =100$.}\label{fig:R-xi}
\end{figure}

On the other hand, both the fluid and effective compression ratios, and thus also the spectral slope, strongly depend on the CR acceleration efficiency. This is evident from Figure~{\ref{fig:R-xi}}, where one can see that, for $\xi_{CR}$ ranging in $\sim 0.03-0.3$,  the slope ranges from $\sim 4.1$ to $\sim 4.6$. This result shows that the inclusion of the magnetic field amplification and of the Alfv\'{e}nic drift in the calculation of the compression factor leads  to quite steep source spectra, with larger $\xi_{CR}$ corresponding to steeper spectra.
Note that within the present setup the fluid compression factor always remains $\lesssim 4$, while in the standard NLDSA theories compression ratios much larger than $4$ are usually found and the CR acceleration efficiency can be well above $\sim 30\%$ (see e.g. \citealt{Blandford-1987PhR...154....1B}; \citealt{Blasi-2002}).
\section{Comparing the predicted proton and helium spectra with data}
\label{sec:results}
Here we assume that SNR shocks accelerate CRs with an efficiency uniformly distributed in the range $\xi_{CR} \sim \, 0.03-0.3$, which implies, as shown in Section~\ref{sec:method}, a dispersion in the CR spectral  slope, $\gamma_{CR}$, in the range $\sim 4.1-4.6$. Formally, this is the slope of the CR spectrum at the shock, and not that of the spectrum of particles escaping the SNR and injected in the ISM.
However, under reasonable assumptions these two spectra are identical (see e.g. \citealt{Gabici-2011MmSAI..82..760G} and references therein).
After escaping SNRs, CRs are believed to propagate diffusively in the Galaxy with a diffusion coefficient $D(R) = D_0 (R/{\rm GV})^{\delta}$  ($R$ is the particle rigidity). The values of $D_0$ and $\delta$ are chosen in order to  fit the observed proton spectrum in the energy range 40 GeV-10 TeV, namely around the spectral hardening at 200-300 GeV. As for helium, we used  the same injection spectral slopes and diffusion coefficient of protons, but we also took into account spallation.
The proton and helium spectra below $\sim 40 \rm GeV$/nucleon are not considered since at these energies both the solar modulation and possible advection effects are important (see e.g \citealt{Aloisio15}), which were not included in our calculation. 

Under these assumptions the proton spectrum ($E$ is the particle energy) can be written as (see e.g \citealt{Berezinskii90}; \citealt{Blasi-Amato-2012JCAP...01..010B})
\begin{equation}\label{eq:p-flux}
f_p(E) = \int_{\xi_{m}}^{\xi_{M}} \frac{R_{SN}}{\pi R_d^2}\frac{H}{2 D(E)}
g_p(E)\frac{d\xi_{CR}}{\xi_M - \xi_m},
\end{equation}
where 
\begin{equation}
g_p(E) \equiv \frac{\xi_{CR} E_{SN}}{I(\gamma_{CR})(mc^2)^2} \left(\frac{E}{mc^2}\right)^{-\gamma_{CR}+2}.
\end{equation}
$I(\gamma_{CR})= \int_{x_0}^{\infty} dx\,x^{2-\gamma_{CR}}\left[\sqrt{1+x^2} -1\right]$ is a normalization factor chosen in such a way that $\int_{E_0}^{\infty} g_p(E)E_k dE = \xi_{CR}E_{SN}$, where $x\equiv E/mc^2$ and $E_k$ is the particle kinetic energy.\\
The helium spectrum is given by
\begin{equation}\label{eq:He-flux}
f_{He}(E) = \int_{\xi_{m}}^{\xi_{M}} \frac{R_{SN}}{\pi R_d^2}\frac{H}{2 D(E)}
g_{He}(E) \frac{1}{1+ \frac{h\,H\,n_d\,c\,\sigma_{sp}}{D(E)}}\frac{d\xi_{CR}}{\xi_M - \xi_m},
\end{equation}
where
\begin{equation}
g_{He}(E) \equiv \eta_{He} \frac{\xi_{CR} E_{SN}}{I(\gamma_{CR})(mc^2)^2} \left(\frac{E}{mc^2}\right)^{-\gamma_{CR}+2}.
\end{equation}
Here $R_{SN} \approx 1/30 ~\rm yr$ is the SN explosion rate in the Galaxy, $R_d \approx 15 ~\rm kpc$ is the Galactic disc radius, $h \approx 250 ~\rm pc$ is the Galactic disc height, $H \approx 4 ~\rm kpc$ is the Galactic halo size, $n_d \approx 5 ~\rm cm^{-3}$ is the average gas density in the disc. $\eta_{He}$ is a  factor chosen in such a way to reproduce the correct normalization of the helium spectrum. 
Finally, $\sigma_{sp}$ is the helium spallation cross section (see e.g \citealt{Blasi-Amato-2012JCAP...01..010B}). The CR acceleration efficiency is in the range $\xi_m \approx 0.03$ to $\xi_M \approx 0.3$.
\begin{figure}
	\centering
	\includegraphics[width=\linewidth]{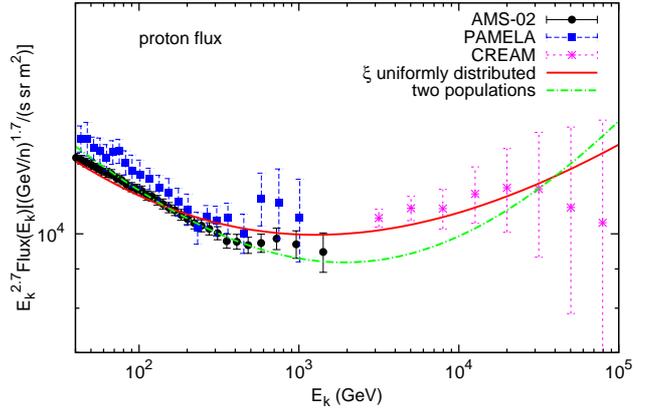}
	\caption{Proton flux compared with the PAMELA, AMS-02 and CREAM data. The plot has been obtained by assuming a spatially independent  CR diffusion coefficient with spectral slope 0.4.}\label{fig:proton-flux}
\end{figure}
\begin{figure}
	\centering
	\includegraphics[width=\linewidth]{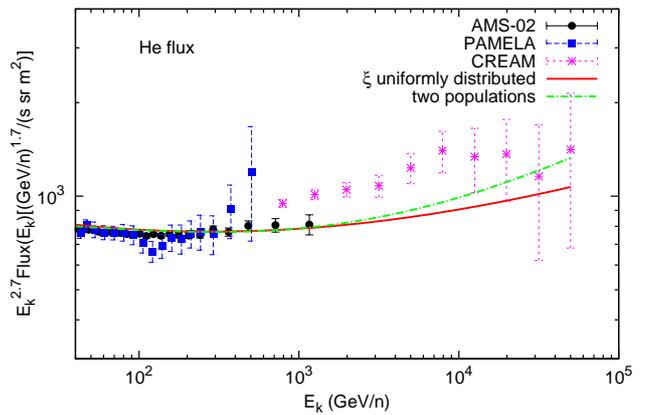}
	\caption{Helium flux compared with the PAMELA, AMS-02 and CREAM data. The plot has been obtained with the  same diffusion coefficient used for the proton flux.}\label{fig:He-flux}
\end{figure}
In Figure~{\ref{fig:proton-flux}} we show the proton flux as computed from Equation~{\ref{eq:p-flux}} (red line) compared with the data by PAMELA (\citealt{Adriani2011}), by AMS-02 (\citealt{Aguilar-2015}) and by CREAM (\citealt{Cream-2011}). The plot has been obtained with the diffusion coefficient parameters: $D_0 \sim 8\times 10^{28} \rm cm^2/s$ and $\delta \sim 0.4$. The slope found for the diffusion coefficient is well within the observational constraints, namely $\delta \sim 0.3-0.6$. With this diffusion coefficient, the grammage traversed by CRs, namely $X= n_d\,h\,H\,c\,m_p/D(R)$, is $\approx 11\rm g/cm^2$ at 10 GeV/n (see e.g. \citealt{Blasi-2013A&ARv..21...70B}). 

On the same Figure we also show the proton flux (green line) computed in the case of two distinct populations of CR sources, one with $\xi_{CR} \approx 0.03$ and the other with $\xi_{CR} \approx 0.3$. Also this plot has been obtained with a diffusion coefficient slope of $\delta = 0.4$, while the explosion rate of the population with the largest acceleration efficiency  has been taken to be $\sim 3$ times smaller than that with the smallest efficiency. Notice that taking into account  such  scenario could be motivated by a different behavior  of type I and II supernovae in the acceleration of CRs (see e.g \citealt{Zatsepin-2006}). In Fig.~{\ref{fig:He-flux}} we show the same as in Fig.~\ref{fig:proton-flux} for the helium flux.

Note that the dispersion in the CR acceleration efficiency, and the consequent dispersion in the CR spectral slope, naturally leads to  a spectral hardening in the proton spectrum at $\lesssim \rm TeV$ energies and, overall, to a good agreement with the data. A similar hardening is found also in the helium spectrum. Moreover, in agreement with observations, this feature is less prominent in the helium spectrum compared to the proton spectrum and the helium spectrum is found to be  harder than the one of protons. This is due to spallation, which hardens the spectrum,  especially at lower energies (see also \citealt{Blasi-Amato-2012JCAP...01..010B}). In the case of two distinct populations with different acceleration efficiencies the spectral hardening is also well reproduced, both in the proton and helium spectrum. However in this case the spectral feature appears to be sharper (see e.g \citealt{Genolini-2017})  than in the case of uniformly distributed efficiency.

Finally, when comparing our results with data, one has to keep in mind that the AMS-02 and CREAM data for helium at $\sim 1$ TeV/nucleon differs by $\sim 20-30$ \%, making it impossible to obtain an equally accurate fit to both data sets.

\section{Conclusions} 
\label{sec:conc}
The magnetic field amplification at SNR shocks, which is thought to be necessary in order to accelerate CRs up to  PeV energies, may also act as a feedback process which limits the maximum achievable CR acceleration efficiency to $\sim 30\%$, thus keeping the overall shock modification modest. Moreover, together with the Alfv\'{e}nic drift, the magnetic field amplification leads to quite steep CR source spectra (spectral slope in momentum$\sim 4.1-4.6$), in agreement with the CR spectra in SNRs inferred from $\gamma-$ray observations (\citealt{Caprioli-2012JCAP...07..038C} and references therein).\\

In this paper we studied the acceleration of CRs at SNR shocks under the following realistic assumptions: 
\begin{enumerate}
\item{the CR acceleration efficiency is may vary within the range $ \sim \,0.03-0.3$;} 
\item{the shock modification induced by the CR pressure at the shock is modest;}
\item{the magnetic field is significantly amplified by CR streaming instability and the Alfv\'{e}n speed (computed in the amplified field) upstream of the shock is enhanced accordingly.} 
\end{enumerate}
We showed that the dispersion in the CR acceleration efficiency produces a dispersion in the shock compression factor. This in turn results is a dispersion in the CR spectral slope, with steeper spectra corresponding to larger acceleration efficiencies.

This result has then be used to demonstrate that, by assuming a diffusive propagation of CRs in the Galaxy with a spatial independent diffusion coefficient, the above mentioned dispersion in the  slope of the injection spectrum can account in a quite natural way for the spectral hardening  found in the proton and helium  spectrum in the energy range $\sim 200-300$ GeV/nucleon.
Moreover, in agreement with observations, because of spallation the helium spectrum is found to be harder than the proton spectrum (even if their injection spectra are identical) and the helium spectral hardening is  less prominent than that of protons.

\section*{Acknowledgments}
SR acknowledges support from the region \^Ile-de-France under the DIM-ACAV programme. SG acknowledges support from the Observatory of Paris (Action Fédératrice CTA). This work has been financially supported by the Programme National Hautes Energies (PNHE) funded by CNRS/INSU-IN2P3, CEA and CNES, France.



\bibliographystyle{mnras}
\bibliography{biblio} 

%
%



\appendix




\bsp	
\label{lastpage}
\end{document}